\documentclass[aps,prc,letterpaper,11pt,twoside,tightenlines,nofootinbib,showpacs,preprint]{revtex4-1}
\usepackage{graphicx}
\usepackage[sort&compress]{natbib}
\usepackage{subfigure}
\usepackage{amsmath}
\usepackage{amsfonts}
\usepackage{cancel}
\usepackage{amssymb}
\usepackage{hyperref}
\begin{document}
\arraycolsep1.5pt
\newcommand{\Ima}{\textrm{Im}}
\newcommand{\Rea}{\textrm{Re}}
\newcommand{\mev}{\textrm{ MeV}}
\newcommand{\be}{\begin{equation}}
\newcommand{\ee}{\end{equation}}
\newcommand{\ba}{\begin{eqnarray}}
\newcommand{\ea}{\end{eqnarray}}
\newcommand{\gev}{\textrm{ GeV}}
\newcommand{\nn}{{\nonumber}}
\newcommand{\dtres}{d^{\hspace{0.1mm} 3}\hspace{-0.5mm}}
\newcommand{\rts}{ \sqrt s}
\newcommand{\non}{\nonumber \\[2mm]}

\title{Isospin breaking and $f_0(980)$-$a_0(980)$ mixing in the $\eta(1405) \to \pi^{0} 
f_0(980)$ reaction}

\author{F. Aceti$^{1}$, W. H. Liang$^{2}$, E. Oset$^{1}$, J. J. Wu$^{3}$ and B. S. Zou$^{4,5}$}
\affiliation{
$^{1}$Departamento de F\'{\i}sica Te\'orica and IFIC, Centro Mixto Universidad de 
Valencia-CSIC,
Institutos de Investigaci\'on de Paterna, Aptdo. 22085, 46071 Valencia,
Spain\\
\\
$^{2}$Physics Department, Guangxi Normal University, Guilin 541004, China\\
\\
$^{3}$Physics Division, Argonne National Laboratory, Argonne, Illinois 60439, USA\\
\\
$^{4}$State Key Laboratory of Theoretical 
Physics, Institute of Theoretical Physics, Chinese Academy of Sciences, Beijing 100190, 
China\\
$^{5}$Theoretical Physics Center for Science Facilities, Institute of High Energy 
Physics, Chinese Academy of Sciences, Beijing 100049, China
}

\date{\today}

\begin{abstract}
We make a theoretical study of the $\eta(1405) \to \pi^{0} f_0(980)$ and
$\eta(1405) \to \pi^{0} a_0(980)$ reactions with an aim to determine the isospin violation 
and the mixing of the $f_0(980)$ and $a_0(980)$ resonances. We make use of the chiral 
unitary approach where these two resonances appear as composite states of two mesons, 
dynamically generated by the meson-meson interaction provided by chiral Lagrangians. We 
obtain a very narrow shape for the $f_0(980)$ production in agreement with a BES 
experiment. As to the amount of isospin violation, or $f_0(980)$ and $a_0(980)$ mixing, assuming constant vertices for the primary $\eta(1405)\rightarrow \pi^{0}K\bar{K}$ and $\eta(1405)\rightarrow \pi^{0}\pi^{0}\eta$ production,
we find  results which are much smaller than found in the recent experimental BES 
paper, but consistent with results found in two other related 
BES experiments. We have tried to understand this anomaly by assuming an I=1 mixture in 
the $\eta(1405)$ wave function, but this leads to a much bigger width of the $f_0(980)$ 
mass distribution than observed experimentally. The problem is solved by using the primary production driven by  $\eta' \to K^* \bar K$ 
followed by $K^* \to K \pi$, which induces an extra singularity in the loop functions 
needed to produce the $f_0(980)$ and $a_0(980)$ resonances. Improving upon earlier work 
along the same lines, and using the chiral unitary approach, we can now predict absolute 
values for the ratio $\Gamma(\pi^0, \pi^+ \pi^-)/\Gamma(\pi^0, \pi^0 \eta)$ which are in 
fair agreement with experiment. We also show  that the same results hold if we had the $\eta(1475)$ resonance or a mixture of these two states, as seems to be the case in the BES experiment.
\end{abstract}

\maketitle

\section{Introduction}
\label{Intro}
In a recent paper the BES team has reported an unusually large isospin violation in the 
decay of the $\eta(1405) \to \pi^{0} f_0(980)$ compared to the
$\eta(1405) \to \pi^{0} a_0(980)$ reaction \cite{etabes}. The $\eta(1405)$ being an isospin 
$I=0$ object can decay naturally to $\pi^{0} a_0(980)$, but the decay into $\pi^{0} f_0(980)$ 
violates isospin. A mixture of the $f_0(980)$ and $a_0(980)$ is unavoidable because 
isospin is broken in meson rescattering due to the different masses of the charged and 
neutral kaons, as was early discussed in \cite{achasov}. More recently the subject has 
been thoroughly discussed in \cite{wulong, ramonetchris} suggesting the study of the $J/\psi \to 
\phi \pi^{0} \eta$ reaction as a test for it. This reaction has been done at BES 
\cite{besphiaf},  where one finds a narrow signal for the $J/\psi \to \phi \pi^{0} \eta$ of 
the order of the difference of kaon masses, as predicted \cite{achasov, wulong, ramonetchris}, 
with and intensity of about half per cent with respect to the one of the $J/\psi \to \phi 
\pi \pi$ in the $f_0(980)$ peak of the  $\pi \pi$ mass distribution. 
Very recently, this reaction has been studied theoretically in \cite{luisnew} using the chiral unitary approach, as in \cite{ramonetchris}, showing that one not only gets the shape of the experiment but also the absolute rate.
Following the suggestion of \cite{wujj2}, the same experimental work of \cite{besphiaf} also reports on the $\chi_{c1} \to \pi^{0} \pi \pi$ in the region of the 
$f_0(980)$ peak of the  $\pi \pi$ mass distribution, and once again finds a narrow 
signal, with an intensity with respect to $\chi_{c1} \to \pi^0 \pi^0 \eta$ in the 
$a_0(980)$ region of the  $\pi^0 \eta$ mass distribution of the order of also half per 
cent. These numbers are within expected values for isospin violation and the narrowness 
of the isospin forbidden signal is tied to the mass difference between charged and 
neutral kaons, reflecting that the isospin violation is tied to the difference of the 
loop functions of intermediate kaons in the rescattering of mesons that leads both to the 
$f_0(980)$ and $a_0(980)$ resonances. This provides support \cite{ramonetchris} to the 
chiral dynamical picture of these resonances 
\cite{npa,ramonet,kaiser,markushin,juanito,rios,nebreda}, which appear as composite 
states of meson-meson, dynamically generated by the interaction of mesons  provided by 
the chiral Lagrangians \cite{weinberg,gasser}.

 With this earlier experimental work, the recent work on the $\eta(1405) \to \pi^0 
f_0(980)$ and $\eta(1405) \to \pi^0 a_0(980)$ reactions \cite{etabes} has brought a 
surprise. The signal for the isospin violating channel $\eta(1405) \to \pi^0 f_0(980)$ is 
also very narrow, in agreement with previous findings in analogous reactions, but the 
reported ratio of the partial decay widths of the two channels is abnormally large, $18 \%$ 
for  $\eta(1405) \to \pi^0 \pi^{+} \pi^{-}$ to $\eta(1405) \to \pi^0 a_0(980)$, or summing 
the $\pi^0 \pi^0$ channel to the
$\pi^{+} \pi^{-}$, a ratio of $27\%$ for the ratio of rates of  $\eta(1405) \to \pi^0 
f_0(980)$ to $\eta(1405) \to \pi^0 a_0(980)$. One anticipates difficulties in a 
theoretical description of such a large rate, unless the same $\eta(1405)$ state already 
contains a large mixture of $I=0$ and  $I=1$, in which case the rate of production of the 
$f_0(980)$ final state would be largely enhanced. However, in this case, the signal of 
the $f_0(980)$ would not be due to the difference of the kaon masses and the production 
of the $f_0(980)$ would proceed unhindered, showing the natural width of the $f_0(980)$ 
of about 50 MeV instead of the 9 MeV observed in the BES experiment \cite{etabes}.

In \cite{wuzou} a particular mechanism was proposed, consisting in the $\eta(1405)$ decay into $K^{*}\bar{K}$, the posterior $K^{*}$ decay into $\pi^{0}K$ and the rescattering of the $K\bar{K}$ to produce either the $f_{0}(980)$ and the $a_{0}(980)$ resonances. This leads technically to a triangular loop diagram that has two cuts (singularities in the integrand), which make it different from the standard $G$ loop function from $K\bar{K}$, with only the $K\bar{K}$ on shell singularity. This latter $G$ function would appear should the $\eta'\rightarrow\pi^{0}K\bar{K}$ vertex be a contact term, or if it was coming from diagrams where an internal propagator is far off shell (contact like vertex).

In the present work we shall make first a thorough discussion of the issue assuming a contact (or contact like) $\eta'\rightarrow\pi^{0}K\bar{K}$  
vertex. Under this assumption one can make a quite model independent study, and the conclusion is that the results obtained are in line with those of other reactions, like the $J/\psi\rightarrow\phi\pi^{0}\eta(\pi\pi)$.

A second part is devoted to the explicit study of the triangular mechanism of \cite{wuzou} which is quite unique to the present reaction. Using the chiral unitary approach we shall see that we are able to evaluate the ratio for isospin violation rather reliably, beyond the reach of \cite{wuzou} where the ratio of widths for $\eta'\rightarrow\pi^{0}\pi^{+}\pi^{-}$ and $\eta'\rightarrow\pi^{0}\pi^{0}\eta$ was dependent on an unknown cut off. We find that this ratio is sizeably increased with respect to the standard approach, in the line of the claims of \cite{wuzou}. We also show that the consideration of extra mechanism driven by primary $\pi^{0}\pi^{0}\eta$ production and rescattering can further increase a bit that ratio such that a good agreement with experiment is found at the end. We emphasize that the concept of $f_{0}-a_{0}$ mixing is not very appropriate since the apparent mixing is so different in different reactions. We rather prefer to talk in terms of isospin violation, magnified due to the proximity of the $f_{0}$ and $a_{0}$ resonances, but which is very much tied to different reactions. The ability of the chiral unitary approach to provide a fair description of all these processes certainly gives support to this method and the underlying consequence in this case, that the $f_{0}$ and $a_{0}$ resonances are basically molecular states of meson-meson, mostly $K\bar{K}$ in both cases.

In what follows we will assume that we have the $\eta(1405)$ decay, while in the BES experiment a mixture if the $\eta(1405)$ and $\eta(1475)$ is present. We also evaluate decay rates for the $\eta(1475)$ and find that the results are basically the same, independently of whether we have either resonance or a mixture of the two.

\section{Formalism}
\label{formalism}
The starting point in the following discussion is the acceptance that the $f_0(980)$ and 
$a_0(980)$ qualify as composite meson-meson states which are dynamically generated  by 
the meson-meson interaction provided by the chiral Lagrangians. The Schr\"{o}dinger 
equation is solved using the kernel (potential) from the chiral Lagrangians, which 
provide a scattering amplitude from where the $f_0(980)$ and $a_0(980)$ emerge as poles 
in the complex plane. In practice, the Bethe-Salpeter equation in coupled channels is used, accounting for dynamical 
and relativistic effects. The basic building blocks are 
$\pi\pi$ and $K\bar{K}$ for the $f_0(980)$ and $\pi\eta$ and $K\bar{K}$ for the 
$a_0(980)$ \cite{npa, ramonet, kaiser, markushin, juanito, rios, nebreda}. Once this is accepted, 
the next step is that, consistently with this picture, these resonances do not couple 
directly to external sources. It is the constituents, pairs of mesons, that couple 
directly to these sources and, upon unitarization (multiple scattering of these mesons) 
the resonances are formed. According to this picture, a series of reactions where these 
resonances are formed were studied and, with no extra parameters than those needed in the 
study of meson-meson scattering, predictions were
made for cross sections or other observables in these reactions. Examples of it are the 
reactions $\phi\rightarrow \pi^0\pi^0\gamma$, $\pi^0\eta\gamma$ \cite{luispalomar}, the 
$J/\psi\rightarrow \phi (\omega) f_0$  
\cite{ollermeissner,palomarchiang,lahde,chrisjpsi}, the $J/\psi\rightarrow 
p\bar{p}\pi\pi$ reaction \cite{li} or the photoproduction of $f_0(980)$ on nucleons 
\cite{marco2}.

The success in the study of these reactions gives strong support to the basic idea that 
we adopt here concerning these resonances.

%************************************************************************************************************************************************
\subsection{Standard formalism assuming local primary $\eta(1405)\rightarrow\pi^{0}PP$ vertices}
\begin{figure}[htpb]
\centering
\includegraphics[scale=0.6]{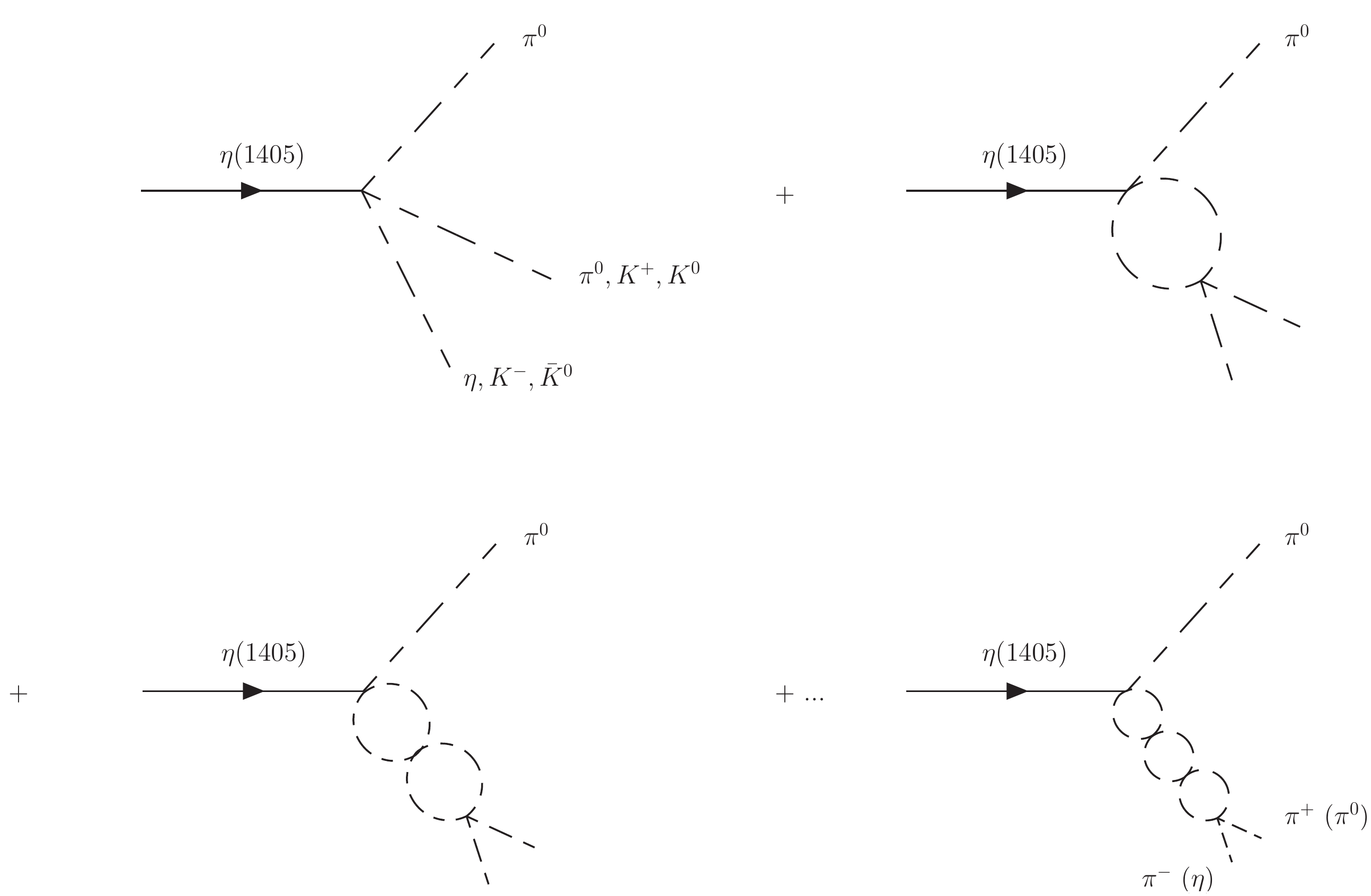}
\caption{Diagrammatic representation of the $\pi^{0}\pi^{+}\pi^{-}$, $\pi^{0}\pi^{0}\eta$ 
production in the $\eta(1405)$ decay.}
\label{fig:diag}
\end{figure}

After this introductory discussion, let us begin the first point where we shall assume that the first step consists of $\eta(1405)\rightarrow\pi^{0}PP$ ($P$ for pseudoscalar) well described by contact (or contact like) vertices. We also accept that the $\eta(1405)$ is an isospin zero state. Then, the mechanism for production of either $\pi^{+}\pi^{-}$ or $\pi^0\eta$ in the final state, together with an extra $\pi^0$, is given by Fig. \ref{fig:diag}.

Implicit in the picture of Fig. \ref{fig:diag} is the fact that the $\pi^0$ of the upper line 
has an energy, when the other pair of mesons produce the $f_0(980)$ or $a_0(980)$, which 
does not match with the energy of the other mesons (in the case of $\pi^0\eta$ 
production) to produce the $f_0(980)$ or $a_0(980)$ resonances. In the case of 
$\pi^0\pi^{+}\pi^{-}$, the $\pi^0$ would not produce either the $f_0(980)$, that has zero 
charge, nor the $a_0(980)$ which does not couple to two pions. But even if it had, it would not play a role in the reaction as we shall discuss below.

The pair of interacting mesons in Fig. \ref{fig:diag} will have $I=1$ if we invoke exact $I=0$ 
for the $\eta(1405)$. Then the $K\bar{K}$ pair appears in the $I=1$ combination
\begin{equation}
\label{eq:kk}
\frac{1}{\sqrt{2}}(K^{+}K^{-}-K^0\bar{K}^0)\ ,
\end{equation}
where we take the convention that $|K^{-}\rangle\equiv -|1/2,-1/2\rangle$ of isospin. 
Should the kaons have the same mass, the loop functions in the figure would be the same 
for charged and neutral kaons and the relative minus sign in Eq. (\ref{eq:kk}) guarantees 
that $\pi^{+}\pi^{-}$ will not be produced, since there is an exact cancellation of the 
$K^{+}K^{-}$ and $K^0\bar{K}^0$ contributions (the $\pi^0\eta\rightarrow\pi^{+}\pi^{-}$ would 
also not proceed). However, when the physical masses are considered, the exact 
cancellation turns into a partial cancellation, leading to an isospin breaking effect 
that we study in detail below.

So far we have only advocated isospin conservation in the  $\eta\rightarrow\pi^0 MM$ 
vertex. Now we can go one step further to put some constraints on the $\pi^0\eta$ primary 
production using arguments of $SU(3)$.

By analogy to the $\eta$ and $\eta'$, which are members of a nonet, with the $\eta$ 
largely an octet and the $\eta'$ basically a singlet, with a small mixing \cite{bramon, 
mathieu, ambrosino}, we can also assume that in the next pair of $\eta$ states, the 
$\eta(1235)$ is largely an octet and the $\eta(1405)$ is mostly a singlet (we shall 
release this constraint later on to quantify uncertainties).

In this case we have to place the interacting meson pair into an octet to produce a 
singlet with the octet of the spectator $\pi^0$. Then, up to an undetermined reduced 
matrix element, the weight of  $K^{+}K^{-}$, $K^0\bar{K}^0$ and $\pi^0\eta$ is determined by 
the $SU(3)$ Clebsch-Gordan coefficients of the $8\otimes8\rightarrow 1$ decomposition, and we have up to a global factor,
\begin{equation}
\label{eq:CGcoeff}
M_{K^{+}K^{-}}=\sqrt{\frac{3}{5}}\ ,\ \ \ \ \ \ \ \  M_{K^0\bar{K}^0}=-\sqrt{\frac{3}{5}}\ ,\ 
\ \ \ \ \ \ \ M_{\pi^0\eta}=\sqrt{\frac{4}{5}}\ .
\end{equation}

Then, the scattering matrix for the production of the final state is given by
\begin{equation}
\label{eq:scatt}
t_f=M_f+\sum_{i=1}^{3}M_iG_iT_{if}\ ,
\end{equation}
where $T_{if}$ is the $5\times 5$ scattering matrix for the channels $K^{+}K^{-}$ (1), 
$K^0\bar{K}^0$ (2), $\pi^0\eta$ (3), $\pi^{+}\pi^{-}$ (4), $\pi^0\pi^0$ (5) and $M_i$ in the 
same basis is given by
\begin{equation}
\label{eq:M}
M_i=A\left(\sqrt{\frac{3}{5}},-\sqrt{\frac{3}{5}},\sqrt{\frac{4}{5}},0,0\right)\ ,
\end{equation}
with $A$ a reduced matrix element.

The $T$ matrix is obtained using the Bethe-Salpeter equation in the five coupled channels
\begin{equation}
\label{eq:bethesalpeter}
T=[1-VG]^{-1}\ V\ ,
\end{equation}
with $V$ taken from \cite{npa} (care is taken to multiply by $1/\sqrt{2}$ the matrix 
elements in the case of $\pi^0\pi^0$ states, thus implementing the unitary normalization 
which is suited for the sum over intermediate states of identical particles).

The G function is the diagonal loop matrix of the propagators of the intermediate 
particles
\begin{equation}
\label{eq:G1}
G(P^{2})=\int\frac{d^4q}{(2\pi)^4}\frac{1}{q^{2}-m_1^{2}+i\epsilon}\frac{1}{(P-q)^{2}-m_2^{2} + 
i\epsilon},
\end{equation}
with $P$ the total four-momentum ($P^{2}=s$) and $m_1, m_2$ the masses of the particles in 
the considered channel. Upon regularization with a cut off one obtains \cite{npa}
\begin{equation}
\label{eq:G}
G(P^{2})=\int_{|\vec{q}|<q_{max}}\frac{d^3q}{(2\pi)^3}\frac{\omega_1+\omega_2}{2\omega_1\omega_2}\frac{1}{(P^{02}-(\omega_1+\omega_2)^{2}+i\epsilon)}\ 
,
\end{equation}
where $\omega_i=\sqrt{\vec{q}\,^{2}+m_i^{2}}$.

By using a cutoff of $q_{max}=900\ MeV$ we obtain a good description of the $f_0(980)$ 
and $a_0(980)$ resonances, as in \cite{npa}.

Note that in Eq. \eqref{eq:scatt} we have two sources of isospin violation. The one due to the $G_i$ functions, which now are different for $K^{+}K^{-}$ and $K^{0}\bar{K}^{0}$, and the $T_{if}$ matrix elements, which are evaluated by means of Eq. \eqref{eq:bethesalpeter} in the charge basis of the states and that also break isospin symmetry because the $G_i$ functions are different for different members of the same isospin multiplets.

Now we would like to restrict the assumption of the $\eta(1405)$ being an $SU(3)$ singlet. Let us accept that it would also have a mixture with an octet. In the case of a pure octet for the $\eta(1405)$ then the interacting pair can belong to the $8$, $10$ and $27$ representations.

Defining  
\begin{equation}
\label{eq:RR}
R=\frac{M(\pi^{0}\eta)}{M(K^{+}K^{-})}\ ,
\end{equation}
we have $R=\sqrt{4/3}$ for  the octet, $R=0$ for the decuplet and $R=-\sqrt{3}$ for the $27$. It is quite unlikely that the $\eta(1405)$ would be a pure octet, and that in this case the interacting pair would couple only to the the $27$ representation, which leads us to values of $R$ preferably positive. Note that with negative values of $R$ (we have seen that this can happen for values around $R\simeq -1.5$) there is  a destructive interference between $\pi^{0}\eta$ and $K\bar{K}$ induced $a_0(980)$ production such that $\pi^{0}a_{0}(980)$ production would disappear in the $\eta(1405)$ decay, which is not the case experimentally \cite{Nakamura:2010zzi}. The order of magnitude for $R$ is determined  with these simple arguments, but we can get help from experiment since we have the ratio \cite{Nakamura:2010zzi,amsler}
\begin{equation}
\label{eq:RRexp}
R_{\Gamma}=\frac{\Gamma(\pi\pi\eta)}{\Gamma(\pi K\bar{K})}=1.09\pm 0.48\ .
\end{equation}
Assuming the ratio to hold for the rates to $\pi^{0}\pi^{0}\eta$ and $\pi^{0}(K^{+}K^{-}+K^{0}\bar{K}^{0})$ we obtain
\begin{equation}
\label{eq:RRexp2}
R_{\Gamma}=\frac{1}{2}\ R^{2}\ \frac{PS(\pi^{0}\pi^{0}\eta)}{PS(\pi^{0} K\bar{K})}\ ,
\end{equation}
where $PS$ stands for the phase space of each final state, which is obtained integrating $\frac{d\Gamma}{dm_{f}}$ of Eq. (\ref{eq:dgamma}) over $m_f$  (taking $\beta=|t_f|=1$). By doing this we obtain 
\begin{equation}
\label{eq:RRexp3}
|R|=0.75\pm 0.17 .
\end{equation}
This result with positive sign would be in agreement with the prediction based  on the assumption of the $\eta(1405)$ being an $SU(3)$ singlet, $R=\sqrt{4/3}=1.15$. Yet, we shall explore the results within the range $R\in[-1,1.2]$.
%************************************************************************************************************************************************

\subsection{Results with the local vertices}
We need to evaluate $\frac{d\Gamma}{dm_f}$ to compare with experiment, where $m_f$ is the 
invariant mass of the final interacting pair ($\pi^{+}\pi^{-}$ and $\pi^{0}\eta$ in our case). 
Since the  meson-meson interaction that leads to the $f_0(980)$ and $a_0(980)$ resonances 
is $s$-wave, there is no angular dependence in the $t_f$ matrix and, since we are 
concerned only around the $m_f=980\ MeV$ region, the magnitude $A$ in Eq. (\ref{eq:M}) 
can be considered constant. In this case we have \cite{nacher}
\begin{equation}
\label{eq:dgamma}
\frac{d\Gamma}{dm_f}=\beta\  p_1\ \tilde{p}_2\ |t_f|^{2}\ ,
\end{equation}
with $\beta$ a constant factor, where $p_1$, $\tilde{p}_2$ are the momentum of the 
spectator $\pi^{0}$ in the $\eta(1405)$ rest frame and the momentum of the interacting pair 
in the rest frame of the pair, respectively
\begin{equation}
\begin{split}
\label{eq:momentums}
&p_1=\frac{\lambda^{1/2}(m_{\eta'}^{2},m^{2}_{\pi^{0}}, m_f^{2})}{2m_{\eta}}\ ,\\
&\tilde{p}_2=\frac{\lambda^{1/2}(m_{f}^{2},m^{2}_{2}, m_3^{2})}{2m_{f}}\ .
\end{split}
\end{equation}
In Eqs. (\ref{eq:momentums}), $\lambda$ is the K\"allen function and $m_2$, $m_3$ the 
masses of the mesons of the interacting pair.

\begin{figure}[ht]
\begin{minipage}[b]{0.7\linewidth}
\centering
\includegraphics[width=\textwidth]{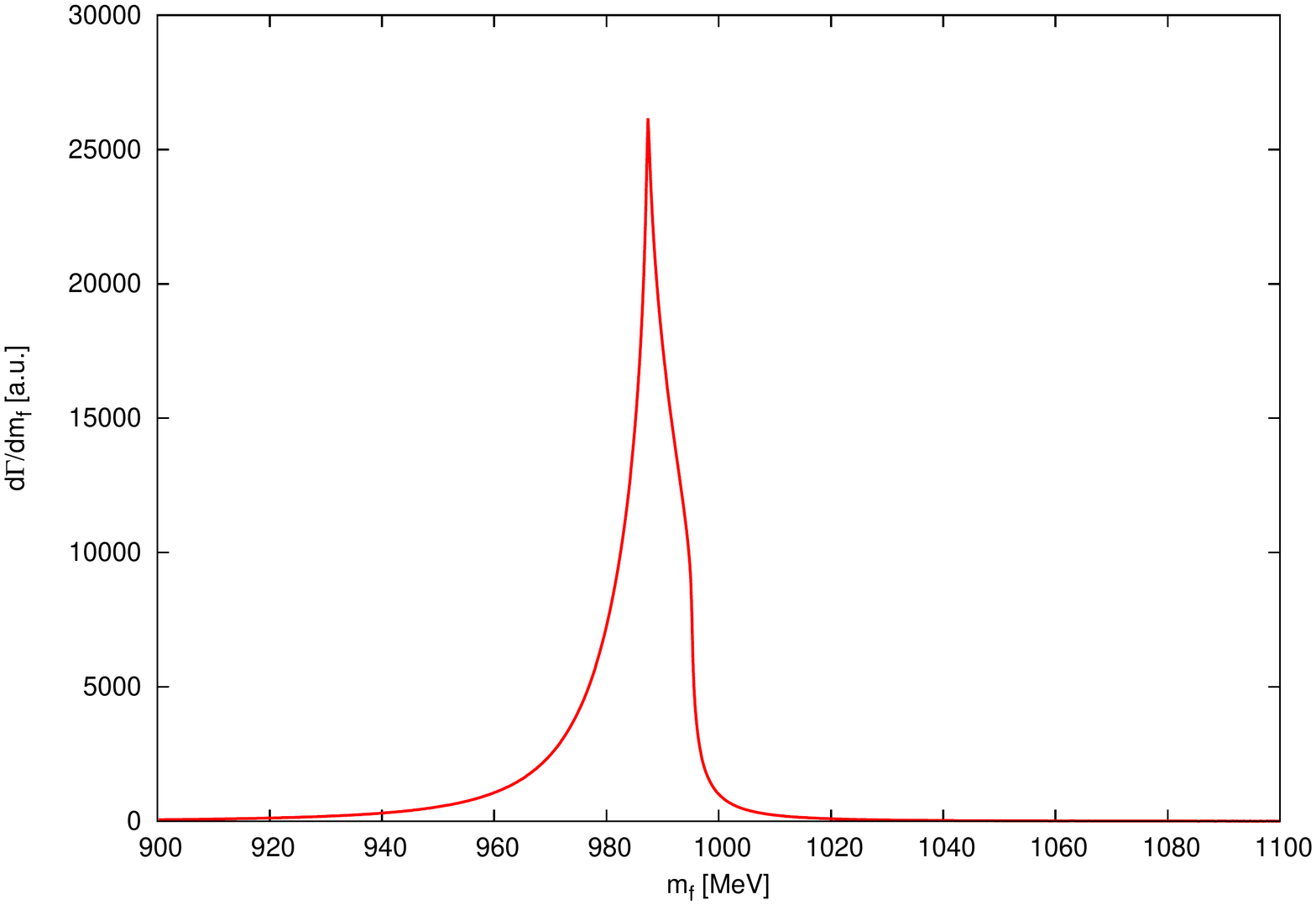}
\caption{$\frac{d\Gamma}{dm_f}$ for $\eta'\rightarrow\pi^{0}\pi^{+}\pi^{-}$ decay in the 
$f_0(980)$ region.}
\label{fig:figure2a}
\end{minipage}
\hspace{0.5cm}
\begin{minipage}[b]{0.7\linewidth}
\centering
\includegraphics[width=\textwidth]{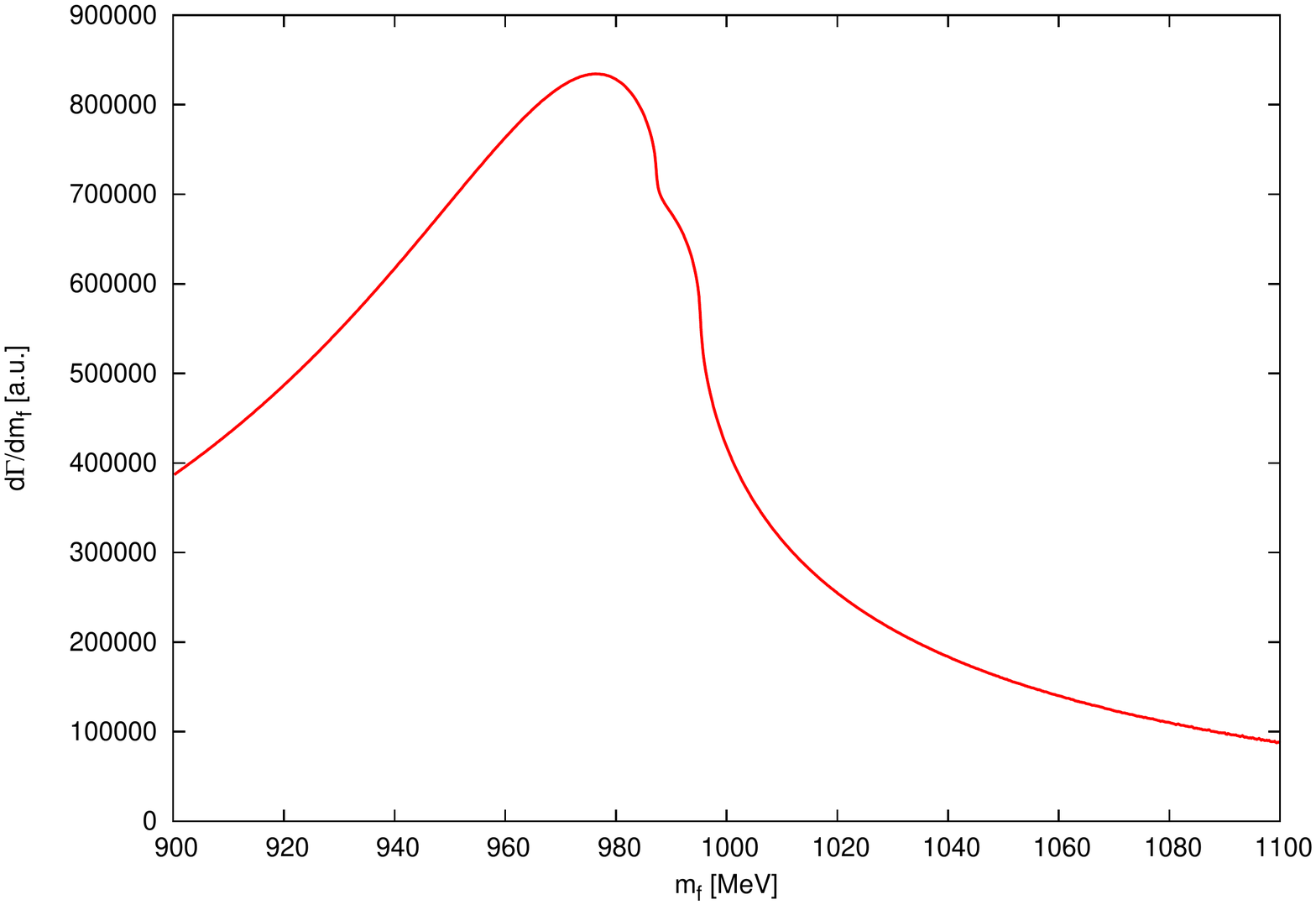}
\caption{$\frac{d\Gamma}{dm_f}$ for $\eta'\rightarrow\pi^{0}\pi^{0}\eta$ decay in the 
$a_0(980)$ region.}
\label{fig:figure2b}
\end{minipage}
\end{figure}

In Figs. \ref{fig:figure2a} and \ref{fig:figure2b},  we plot $\frac{d\Gamma}{dm_f}$ for $f$ equal to $\pi^{+}\pi^{-}$ and 
$\pi^{0}\eta$, taking $A$ from Eq. (\ref{eq:M}) equal $1$. We can rightly say that the unitarization from the meson meson pairs should be implemented in other pairs too. Think for instance of primary production of $\pi^{0}K\bar{K}$ and then $\pi^{0}K$ interaction producing an effective $\eta(1405)K \bar{K} \pi^0$ vertex that will depend on $m(\pi^{0}K)$. After this, the $K\bar{K}$ will interact again to finally produce the $f_{0}$ or $a_{0}$. The isospin, or $SU(3)$ argument used before should also hold, but the coefficient $A$ would now be dependent on $m(\pi^{0}K)$ which also introduces an angular dependence on this coefficient. However, upon projection over $s-$wave, needed to generate the $f_{0}$ or $a_{0}$ resonances, and the selection of a narrow window for $m(K\bar{K})$ around $980\ MeV$, the coefficient $A$ turns again into a constant. Similar arguments can be made with respect to the symmetrization of the two pions in the $\pi^{0}\pi^{0}\eta$ channel.

\begin{figure}[ht]
\begin{minipage}[b]{0.5\linewidth}
\centering
\includegraphics[width=\textwidth]{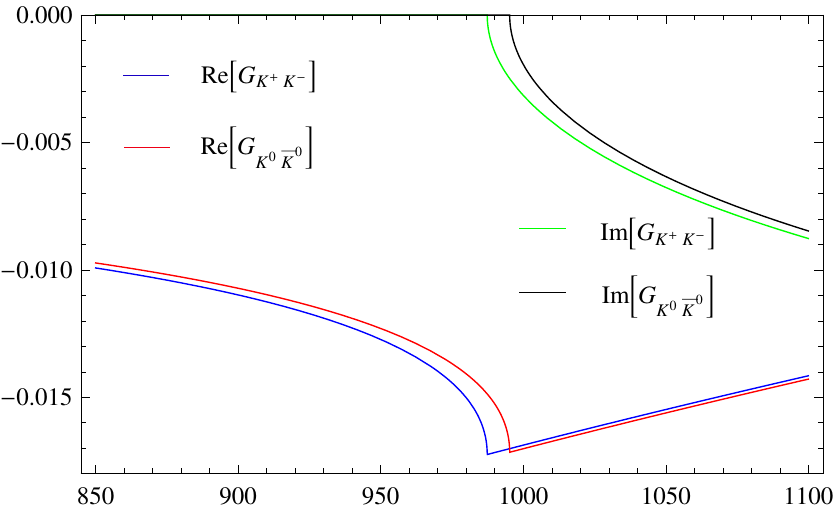}
\caption{Real and imaginary part of $G_{K^{+}K^{-}}$ and $G_{K^{0}\bar{K}^{0}}$ as functions of the energy.}
\label{fig:g}
\end{minipage}
\end{figure}

What we can see in Figs. \ref{fig:figure2a} and \ref{fig:figure2b} is that in the case of the $\pi^{+}\pi^{-}$ production we 
obtain a very narrow peak around $980\ MeV$ like in the experiment \cite{etabes}. The 
width of this peak is about $10\ MeV$, in agreement with experimental observations. As we 
discussed above, the peak appears in the $f_0(980)$ region, in between the thresholds of 
$K^{+} K^{-}$ and $K^{0} \bar{K}^{0}$,  because now $G_{K^{+}K^{-}}-G_{K^{0}\bar{K}^{0}}$ is different 
from zero. However, the difference, which is due to the different kaon masses, is only 
significant in a region of energies around the $K\bar{K}$ thresholds, where 
$\Delta(\sqrt{s})$ is of the order of $m_{K^{+}}-m_{K^{0}}$, see Fig. \ref{fig:g}. 
Away from the thresholds the 
difference of the two $G$ functions due to the mass difference becomes gradually smaller 
and this leads to the peculiar narrow shape of the $f_0(980)$ excitation in the 
$\pi^{+}\pi^{-}$ channel, already anticipated in \cite{achasov,wulong, ramonetchris}.

One should stress here that the shape of Fig. \ref{fig:figure2a} is not the standard one of the $f_{0}(980)$ seen in isospin allowed reactions and the width is tied to the mass difference $m_{K^{+}}-m_{K^{0}}$. This comment is pertinent in view of the comment in \cite{etabes} quoting that ``The measured width of the $f_{0}(980)$ is much narrower than the world average". It is clear that the shape of $\pi^{+}\pi^{-}$ production here is not the shape of the $f_{0}(980)$. 

In Fig. \ref{fig:figure2b} we see the signal for the $a_0(980)$ excitation, which is isospin 
allowed. The width is much larger and the strength at the peak is also much larger. If we 
compare the strength of the peak for $\pi^+ \pi^-$ of $f_0$ and $\pi^0 \eta$ of $a_0$  production, we find that the ratio 
is of the order of $3\%$.  
\begin{figure}[ht]
\begin{minipage}[b]{0.7\linewidth}
\centering
\includegraphics[width=\textwidth]{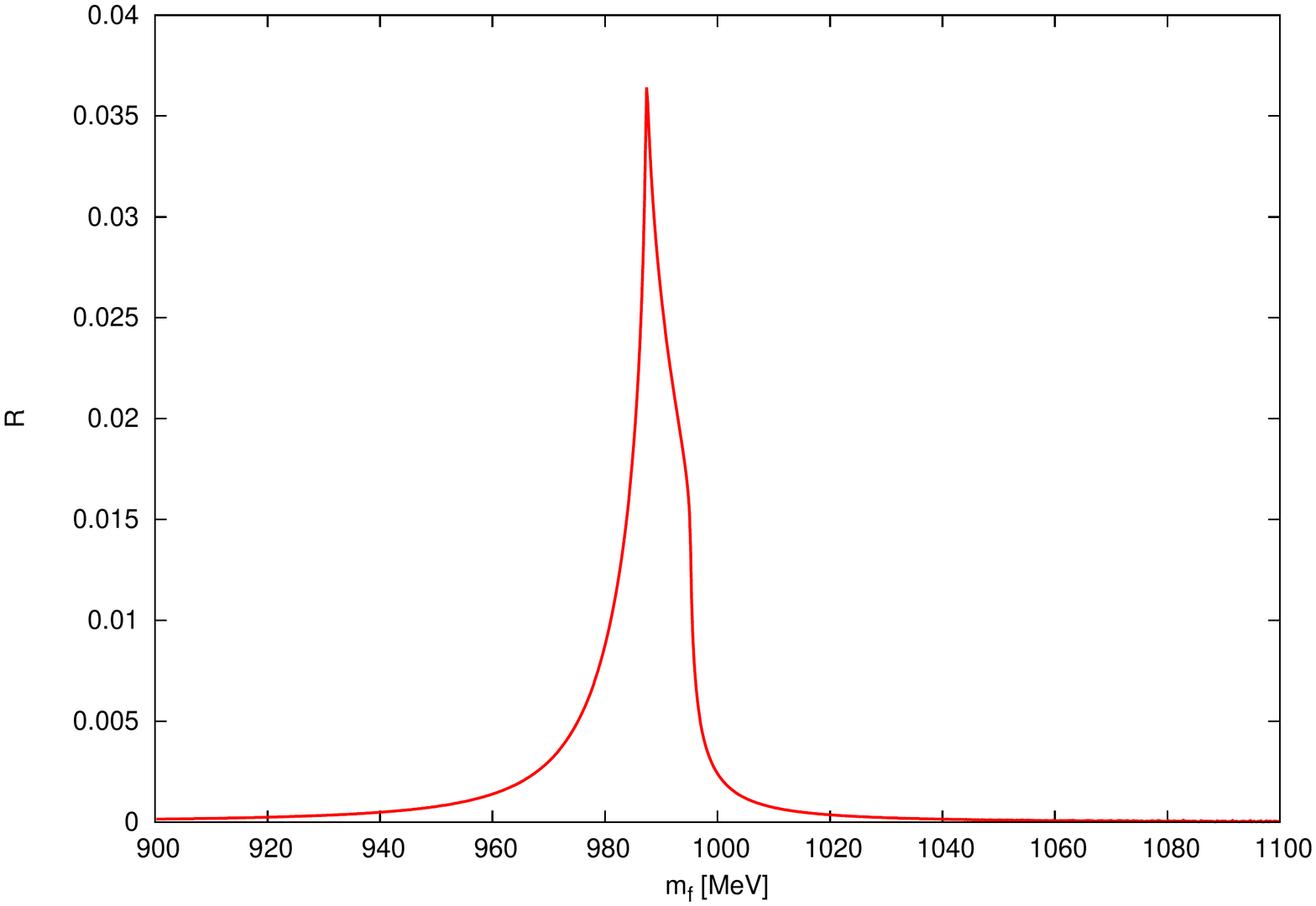}
\caption{Ratio $\left(\frac{d\Gamma}{dm_f}\right)_{\pi^{+}\pi^{-}}/\left(\frac{d\Gamma}{dm_f}\right)_{\pi^{0}\eta}$ as a function of $m_f$.}
\label{fig:ratio1}
\end{minipage}
\end{figure}
However if we integrate the strength over $m_f$ in the region of the peaks for the two cases, we find a smaller ratio
\begin{equation}
\label{eq:ratio}
\frac{\Gamma(\pi^{0},\pi^{+}\pi^{-})}{\Gamma(\pi^{0},\pi^{0}\eta)}=0.015\ ,
\end{equation}
of the order of $1.5\%$, which is along the lines of the $0.6\%$ observed in the two 
reactions $J/\psi\rightarrow\phi\pi^{0}\eta (\pi^{+}\pi^{-})$ or 
$\chi_{c1}\rightarrow\pi^{0}(\pi^{+}\pi^{-})(\pi^{0}\eta)$ \cite{besphiaf}.
In Fig. \ref{fig:ratio1}, we show the ratio of $d\Gamma(\pi^{+}\pi^{-})/d\Gamma(\pi^{0}\eta)$ as a function of the energy. We observe a peculiar structure, where the $K^{+}K^{-}$, $K^{0}\bar{K}^{0}$ thresholds  show up as cusps, as predicted in \cite{wulong,ramonetchris} and also shown in \cite{luisnew}.

We come now to see the uncertainties due to the diversion from the $SU(3)$ hypothesis 
assumed. We allow $R$ of Eq. (\ref{eq:RR}) to vary between $-1$ 
and $1.2$, as discussed in the previous section.
\begin{figure}[ht]
\begin{minipage}[b]{0.7\linewidth}
\centering
\includegraphics[width=\textwidth]{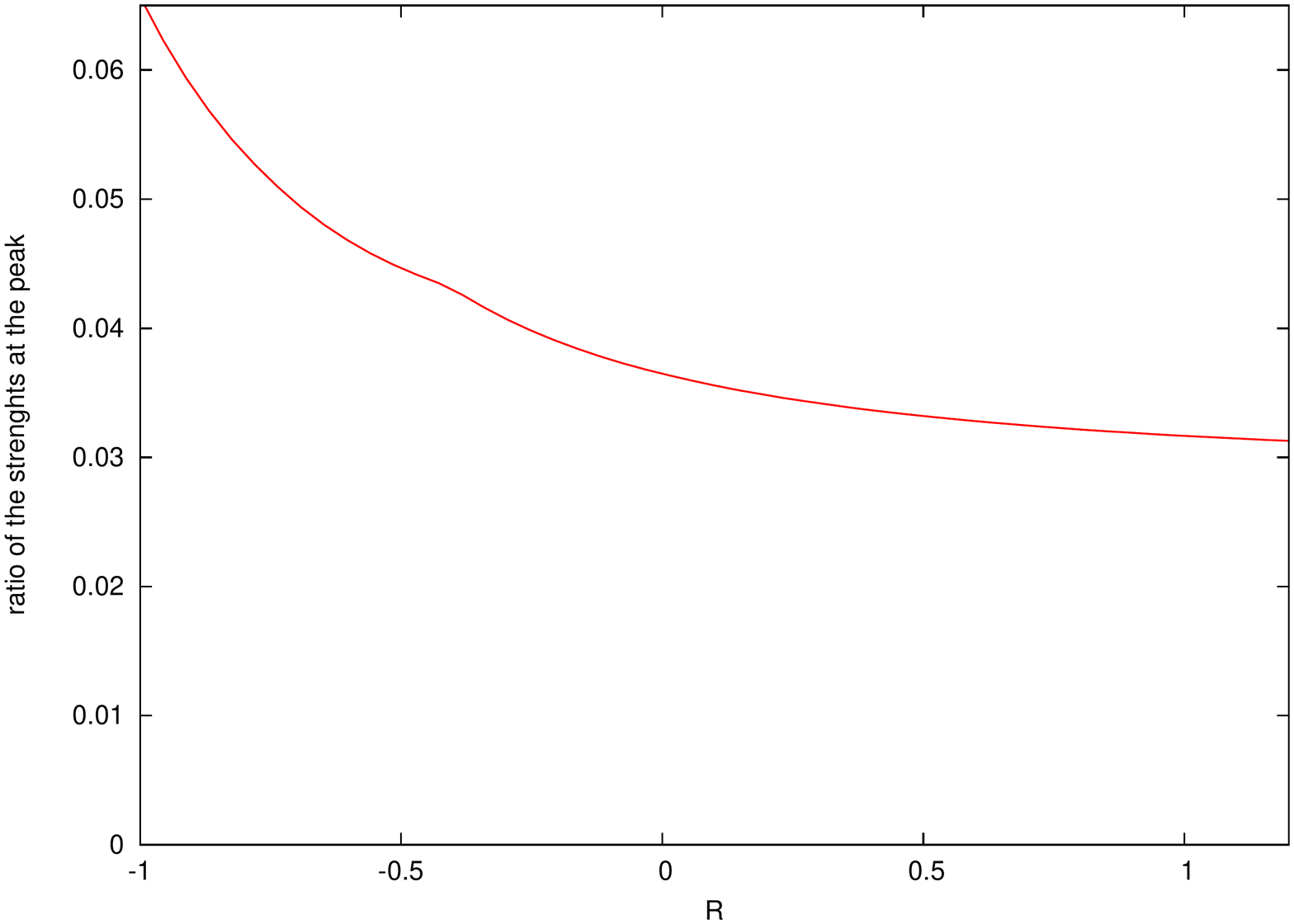}
\caption{Ratio of strengths at the peak as a function of $R$.}
\label{fig:ratio2}
\end{minipage}
\end{figure}
In Fig. \ref{fig:ratio2} we can see that the 
ratio of strengths at the peak of each resonance changes within a factor of two in such a 
large range. In terms of the $m_f$ integrated over the peak, removing background, the range is
\begin{equation}
\label{eq:ratio2}
\frac{\Gamma(\pi^{0},\pi^{+}\pi^{-})}{\Gamma(\pi^{0},\pi^{0}\eta)}\in[0.01-0.04]\ .
\end{equation}
The results are shown in Fig. \ref{fig:ratio3}. At the extreme negative value of $R$, not preferred by the theory, the ratio reaches the value of $0.042$. In the range from $R=0$ (the value implicitly taken in \cite{wuzou}) to $R=1.2$ ($R=1.15$ correspond to the $SU(3)$ singlet for the $\eta(1405)$) the value of the ratio of $\Gamma$'s ranges from $1\%$ to $1.5\%$. Even with this theoretical uncertainty, it is thus clear that we cannot obtain a ratio as big as the $18\%$ reported in the experiment of \cite{etabes}.
\begin{figure}[ht]
\begin{minipage}[b]{0.7\linewidth}
\centering
\includegraphics[width=\textwidth]{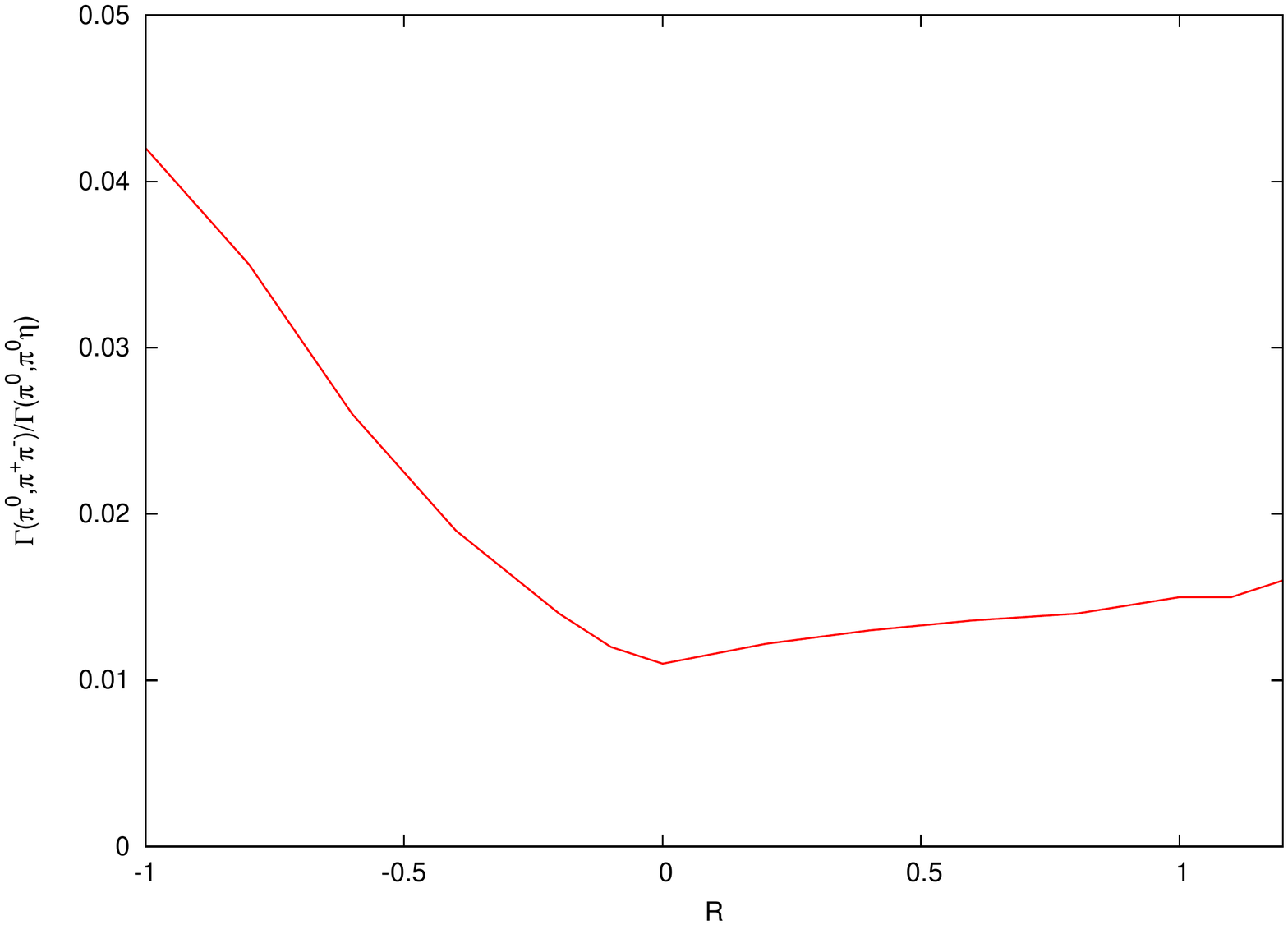}
\caption{Ratio $\frac{\Gamma(\pi^{0},\pi^{+}\pi^{-})}{\Gamma(\pi^{0},\pi^{0}\eta)}$ as a function of $R$.}
\label{fig:ratio3}
\end{minipage}
\end{figure}

There could be a scope, since so far we have always assumed the $\eta(1405)$ to be a pure 
$I=0$ state. Let us assume that we have a mixture of $I=0$ and $I=1$ in that state (the same conclusions would hold if we say instead that there is isospin violation in the production of mesons of the first step, something that is very unusual in chiral theories \cite{ramonetchris}). In 
the case of $I=1$ for the $\eta(1405)$ the interacting meson pair can have $I=0$, which we assume in the $SU(3)$ octet, to magnify the $f_0(980)$ 
production and then the channels are $\pi\pi$ and $K\bar{K}$, but the $\pi\pi$ channel is 
weak in this process and for the exercise that we do can be safely ignored in the production vertices, but not in the $T_{if}$ matrix of Eq. (\ref{eq:scatt}). Then the 
$K\bar{K}$, $I=0$ combination is
\begin{equation}
\label{eq:kk2}
\frac{1}{\sqrt{2}}(K^{+}K^{-}+K^{0}\bar{K}^{0})\ .
\end{equation}

Taking into account the isospin mixture and a different reduced matrix element for $I=0$ pair 
production and putting the product in a coefficient $\alpha$, we have now 
$M_i\rightarrow\tilde{M}_i$, with $\tilde{M}_i$ given by
\begin{equation}
\label{eq:M2}
\tilde{M}_i=A\left((1+\alpha)\sqrt\frac{3}{5}, (\alpha-1)\sqrt\frac{3}{5},\sqrt 
\frac{4}{5}, 0, 0\right)\ .
\end{equation}
We vary the parameter $\alpha$ until we find a ratio 
$\Gamma(\pi^{0},\pi^{+}\pi^{-})/\Gamma(\pi^{0},\pi^{0}\eta)=0.18$. The parameter $\alpha$ has the 
value $0.54$ which implies a massive isospin violation in a physical state. This would be 
difficult to accept in physical terms, but there is one stronger reason to reject this 
solution.
\begin{figure}[ht]
\begin{minipage}[b]{0.7\linewidth}
\centering
\includegraphics[width=\textwidth]{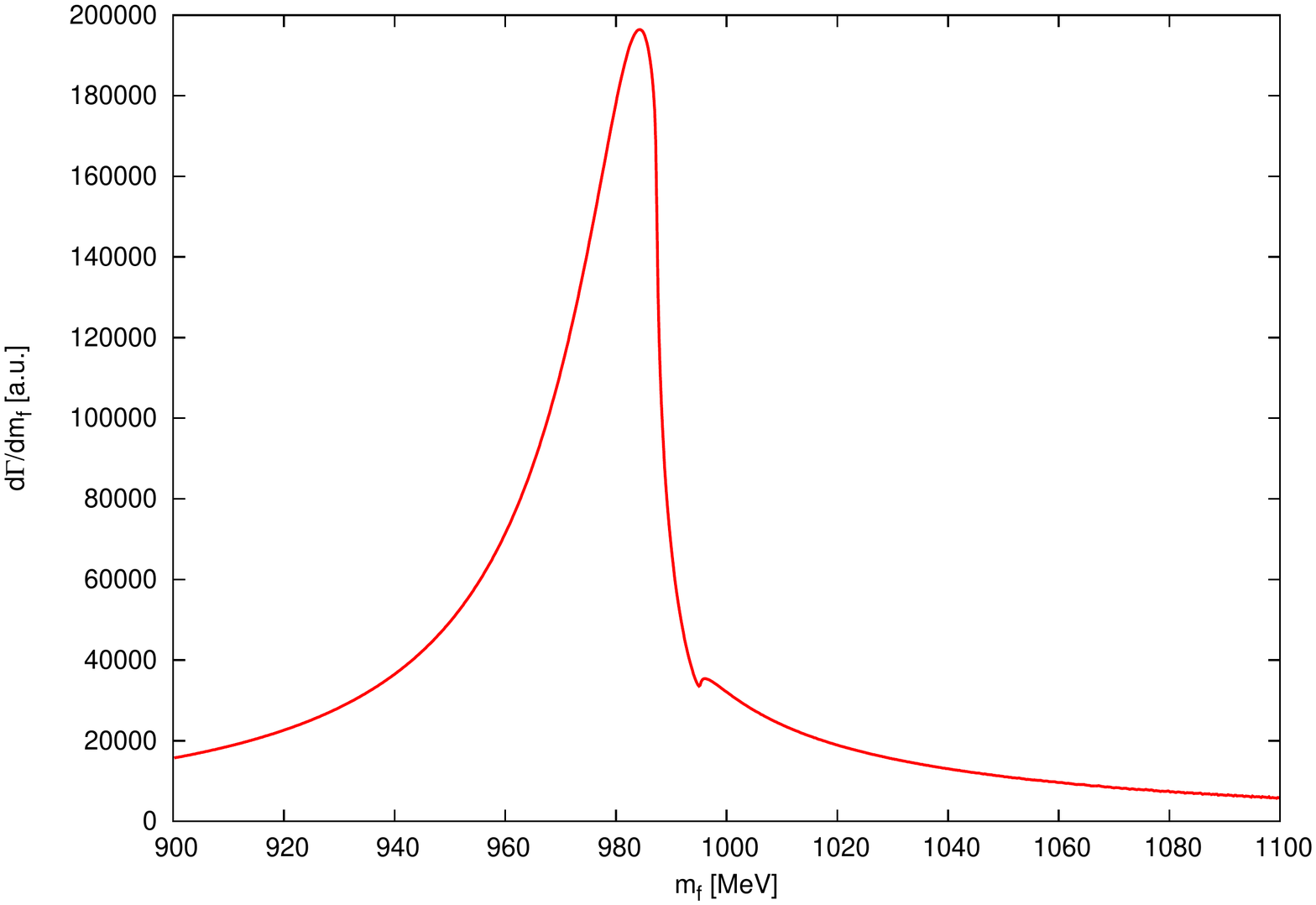}
\caption{$\frac{d\Gamma}{dm_f}$ for $\eta\rightarrow\pi^{0}\pi^{+}\pi^{-}$ decay in the 
$f_0(980)$ region, for $\alpha=0.54$.}
\label{fig:falpha}
\end{minipage}
\hspace{0.5cm}
\begin{minipage}[b]{0.7\linewidth}
\centering
\includegraphics[width=\textwidth]{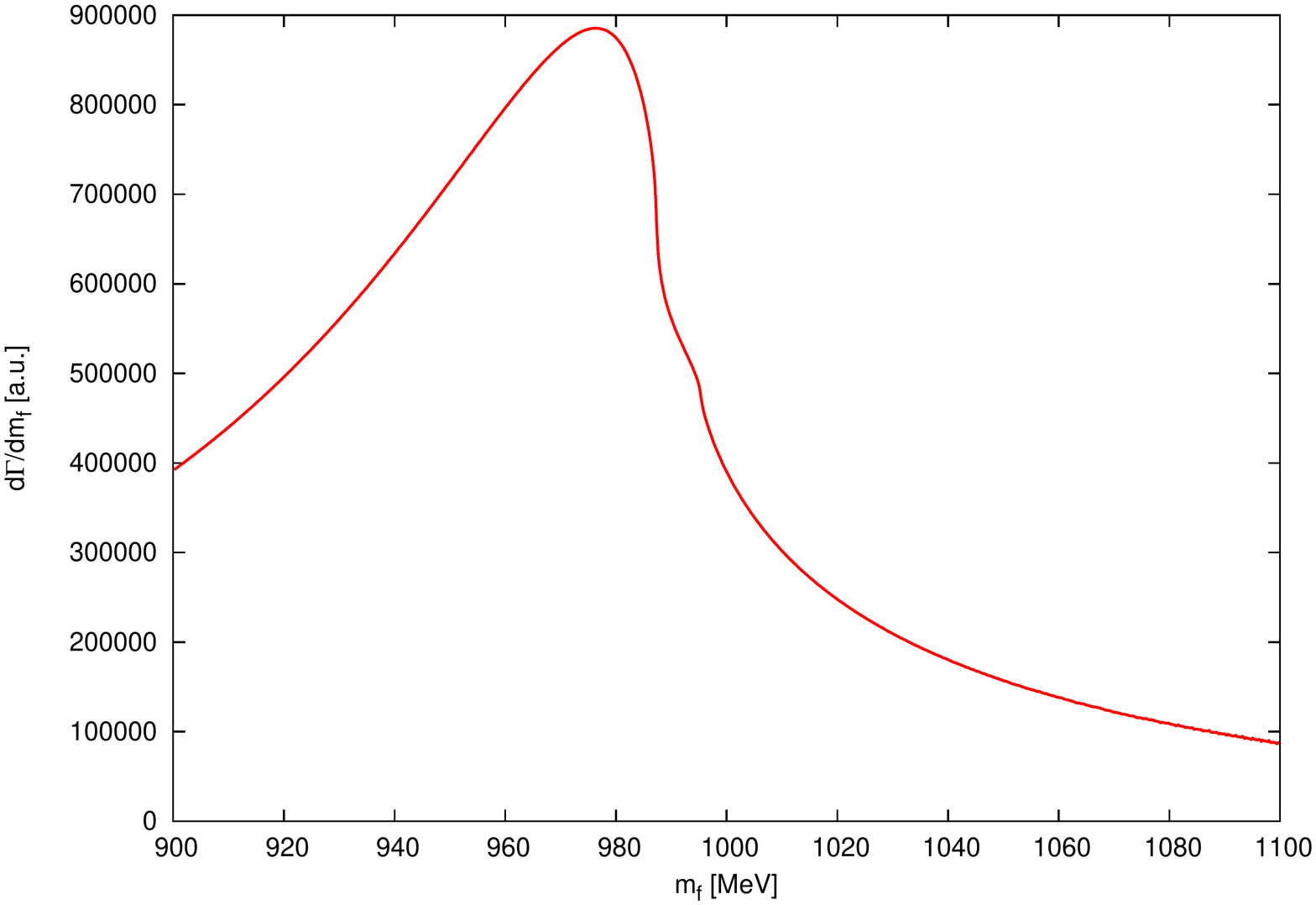}
\caption{$\frac{d\Gamma}{dm_f}$ for $\eta\rightarrow\pi^{0}\pi^{0}\eta$ decay in the 
$a_0(980)$ region, for $\alpha=0.54$.}
\label{fig:aalpha}
\end{minipage}
\end{figure}
Indeed, as seen in Fig. \ref{fig:falpha}, since the $f_0(980)$ production proceeds unhindered because we have an 
$I=0$ pair to begin with, the $f_0(980)$ is produced with its natural width and the combination of Eq. (\ref{eq:M2}) leads to an effective width of about $20\ MeV$, much bigger than the experimentally observed $9\ MeV$ of \cite{etabes}. In Fig. \ref{fig:aalpha} we can see that the $a_{0}(980)$ resonance is also produced in this case with a shape like the ordinary one.
%************************************************************************************************************************************************

\subsection{The primary production vertex with the $K^{*}\bar{K}$ singularity}
\begin{figure}[ht]
\centering
\includegraphics[width=7cm]{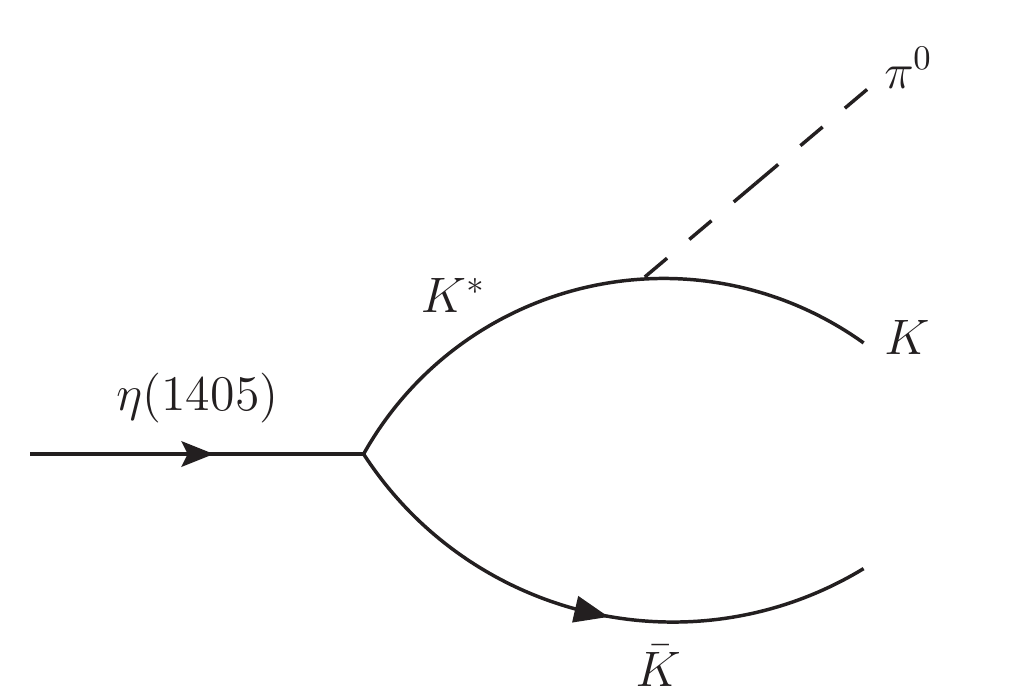}
\caption{Singular mechanism for $\pi^{0}K\bar{K}$ production.}
\label{fig:trian}
\end{figure}

In the former section we showed that it is  not possible to get such a large isospin violation as found in \cite{etabes} assuming a local vertex production. In \cite{wuzou} it was shown that using the $\eta(1405)$ decay mode to $K^{*}\bar{K}$ and the successive decay of $K^{*}$ into $K\pi$ one obtains a mechanism for $K\bar{K}\pi$ production at tree level by means of which one could obtain good agreement with experimental data on this channel. This production mechanism is depicted in Fig. \ref{fig:trian}
\begin{figure}[ht]
\centering
\includegraphics[width=10cm]{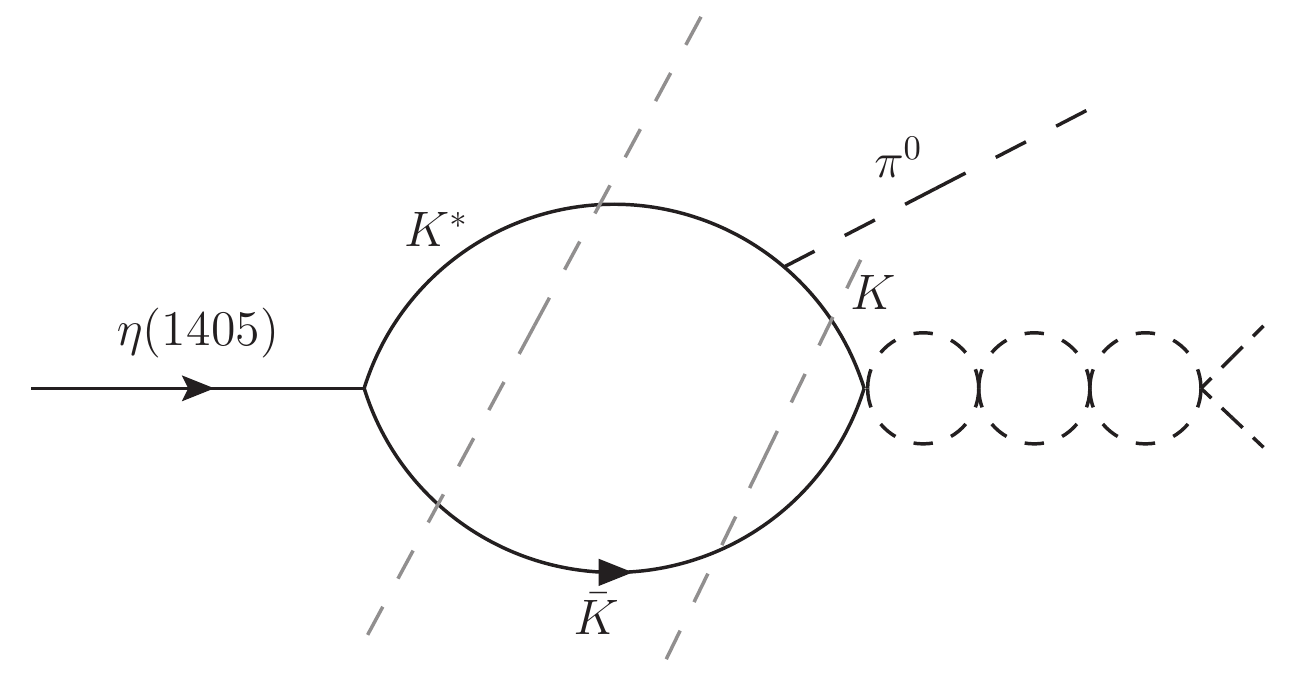}
\caption{Rescattering mechanism for the production of the $f_{0}$ and $a_{0}$.}
\label{fig:trianb}
\end{figure}
After rescattering of the $K\bar{K}$ pair, as shown in Fig. \ref{fig:trianb}, the $f_{0}$ and $a_{0}$ resonances will be produced in our approach. The novelty now is that the first loop depicted in Fig. \ref{fig:trianb} is rather different than the one of the ordinary $G$ function for $K\bar{K}$ propagation shown in the second diagram of Fig. \ref{fig:diag}. The difference is substantial because the structure of the loop function (through dispersion relations) is determined by the singularities (pairs of intermediate particles that can be simultaneously placed on shell in the loop integration). The loop in Fig. \ref{fig:trianb} has two singularity cuts, indicated by the dashed lines, one for the $K^{*}\bar{K}$ on shell and the other one for the $K\bar{K}$ on shell. The kinematics of the two cuts are not too far away, which magnifies the difference in the loop functions in the charged and neutral cases due to the different masses amongst the kaons and the $K^{*}$. 

Note that the situation for $J/\psi\rightarrow\phi f_{0}$ is very different, because even if the highly suppressed $J/\psi\rightarrow K^{*}\bar{K}$ decay would be followed by the $K^{*}\rightarrow\phi K$ vertex, this latter process is kinematically forbidden and then the $K^{*}$ is highly off shell. So, this mechanism for $J/\psi\rightarrow\phi K\bar{K}$ qualifies as a contact term for $\phi K\bar{K}$ production. Then the approach followed in the former section is most appropriate for this case and it is in essence the one followed in \cite{ramonetchris, luisnew}. The experimental ratio for the $J/\psi$ decay widths in this reaction are in line  with the results obtained in the former sections.

On the other hand, the mechanism depicted in Fig. \ref{fig:trian} reminds one of the $\phi\rightarrow\pi^{0}\pi^{0}\gamma$ decay which has the same structure with $\phi\rightarrow K\bar{K}$, the $K$ (or $\bar{K}$) radiating a photon and the resulting $K\bar{K}$ pair interacting to give $\pi^{0}\pi^{0}$ or $\pi^{0}\eta$ (same diagram as Fig. \ref{fig:trianb} substituting the $\pi^{0}$ by $\gamma$ and the $K^{*}$ by $K$). One has there two cuts for $K\bar{K}$ before and after the radiation of the photon. One should then recall that the mechanism outlined above was very successful \cite{marco,mauro,rocaphi,escribano} reproducing the experimental data for $\phi\rightarrow\pi^{0}\pi^{0}\gamma,\pi^{0}\eta\gamma$.
\begin{figure}[ht]
\centering
\includegraphics[width=10cm]{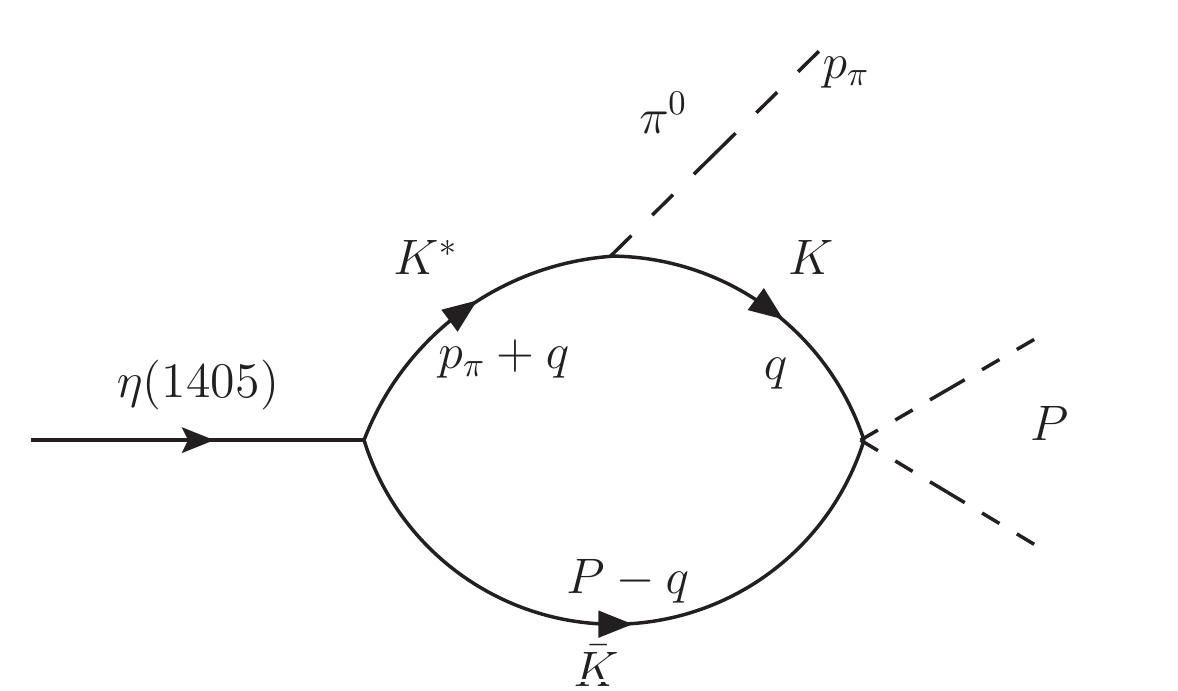}
\caption{Loop for the function $\tilde{G}$.}
\label{fig:diag2}
\end{figure}

Let us proceed to the explicit evaluation of the amplitude for the mechanism of Fig. \ref{fig:trianb}. The loop function is evaluated using the momenta described in Fig. \ref{fig:diag2}. For convenience we make the evaluation in the frame where $\vec{P}=0$ and thus $\vec{p}_{\eta'}=\vec{p}_{\pi}$.

Given the structure of the $V\rightarrow PP$ vertices, $\epsilon^{\mu}(p_1-p_2)_{\mu}$, we obtain
\begin{equation}
\begin{split}
\label{eq:Gtilde}
\tilde{G}(P, p_{\pi}, m_{K}, m_{K^{*}})&=i\int\frac{d^{4}q}{(2\pi)^4}\epsilon^{\mu}(P+p_{\pi}+P-q)_{\mu}\ \epsilon^{\nu}(p_{\pi}-q)_{\nu}\\
&\times\frac{1}{(p_{\pi}+q)^2-m_{K^{*}}^{2}+i\epsilon}\ \frac{1}{q^2-m_{K}^{2}+i\epsilon}\ \frac{1}{(P-q)^2-m_{K}^{2}+i\epsilon}\ .
\end{split}
\end{equation} 
By summing over the polarizations
\begin{equation}
\label{eq:pol}
\sum\epsilon_{\mu}\epsilon_{\nu}\rightarrow -g_{\mu\nu}+\frac{(p_{\pi}+q)_{\mu}(p_{\pi}+q)_{\nu}}{m^{2}_{K^{*}}}\ ,
\end{equation}
we get
\begin{equation}
\begin{split}
\label{eq:Gtilde2}
\tilde{G}(P, p_{\pi}, m_{K}, m_{K^{*}})&=i\int\frac{d^{4}q}{(2\pi)^4}\frac{F_{num}}{(p_{\pi}+q)^2-m_{K^{*}}^{2}+i\epsilon}\ \frac{1}{q^2-m_{K}^{2}+i\epsilon}\\&\times\frac{1}{(P-q)^2-m_{K}^{2}+i\epsilon}\ ,
\end{split}
\end{equation} 
where
\begin{equation}
\begin{split}
\label{eq:fnum}
F_{num}&=-(2P(p_{\pi}-q)+m_{\pi}^2+q^2-2p_{\pi}q)+\frac{(m_{\pi}^2-q^2)}{m_{K^{*}}^{2}}[2P(p_{\pi}+q)+m_{\pi}^2-q^2]\\&=2p_{\eta'}(p_{\pi}-q)+\frac{(m_{\pi}^2-q^2)}{m_{K^{*}}^{2}}[2P(p_{\pi}+q)+m_{\pi}^2+m_{K^{*}}^2-q^2]\ .
\end{split}
\end{equation} 

One technical problem faced in \cite{wuzou} is that the integral of Eq. \eqref{eq:Gtilde2} is highly superficially divergent ($d^{4}q/q^{2}$) and some form factor or cutoffs were used to implement convergence. However, we shall see below that the integral is only logarithmically divergent. When performing the evaluation of  the $\eta'\rightarrow\pi^{0}\pi^{+}\pi^{-}$ amplitude one has the difference of $\tilde{G}$ for the charged $K^{-}K^{+}$ and the neutral one and the results are convergent, but then the ratio to the $\eta'\rightarrow\pi^{0}\pi^{0}\eta$ is tied to an unknown form factor.

Our approach solves naturally the former problem. To see this, recall that in Eq. \eqref{eq:bethesalpeter}, for the scattering, the $G$ function is also formally divergent and is regularized by a cutoff which is fitted to the meson-meson scattering data. The natural choice is to use this cutoff in the  new loop, but this becomes a necessity when one recalls that the results of the chiral unitary approach with the $G$ function implementing a cutoff $\theta(q_{max}-|\vec{q}|)$ in the integration are obtained formally in a Quantum Mechanical formulation starting with a potential (for $s$-waves that we study here)
\begin{equation}
\label{eq:potential}
V(\vec{q},\vec{q}\ ')=v\ \theta(q_{max}-|\vec{q}\ |)\ \theta(q_{max}-|\vec{q}\ '|)\ .
\end{equation} 
Then in Fig. \ref{fig:trianb} the cutoff $\theta(q_{max}-|\vec{q}|)$ appears automatically in the loop function from the first $K\bar{K}\rightarrow PP$ potential in the sum of the diagrams implicit in the figure. Observe that the cutoff is in three-momentum. The $q^{0}$ integration must be done analytically and it is convergent.

The expressions are simplified and equally accurate if we just take the positive energy part of the relativistic $K^{*}$ propagator
\begin{equation}
\label{eq:kstarprop}
\frac{1}{2\omega_{K^{*}}(\vec{p}_{\pi}+\vec{q}\ )}\ \frac{1}{p_{\pi}^{0}+q^{0}-\omega_{K^{*}}(\vec{p}_{\pi}+\vec{q}\ )+i\epsilon}\ ,
\end{equation}
where $\omega_{K^{*}}(\vec{p})=\sqrt{\vec{p}\ ^2+m_{K^{*}}^2}$. Using Cauchy's theorem for the $q^{0}$ integration, we obtain then
\begin{equation}
\begin{split}
\label{eq:Gtilde3}
\tilde{G}(P, p_{\pi}, m_{K}, m_{K^{*}})&=\int_{|\vec{q}|<q_{max}}\frac{d^3q}{(2\pi)^3}\ \frac{1}{2\omega}\ \frac{1}{P^{0}}\ \frac{1}{2\omega_{K^{*}}}\Big[\frac{F_{num}(q^{0}=-\omega)}{P^{0}+2\omega}\ \frac{1}{p_{\pi}^{0}-\omega-\omega_{K^{*}}}\\&+\frac{F_{num}(q^{0}=P^{0}-\omega)}{P^{0}-2\omega+i\epsilon}\ \frac{1}{P^{0}+p_{\pi}^{0}-\omega-\omega_{K^{*}}+i\epsilon}\Big]\ ,
\end{split}
\end{equation}
where $\omega=\sqrt{\vec{q}\ ^2+m_{K}^2}$ and $\omega_{K^{*}}=\sqrt{\vec{q}\ ^2+m_{K^{*}}^2}$. Eq. \eqref{eq:Gtilde3} shows explicitly in the second term the two singularities corresponding to the cuts depicted in Fig. \ref{fig:trianb}. One can show from Eq. \eqref{eq:Gtilde3} that $\tilde{G}$ is only logarithmically divergent. The apparent two extra power of $q$ introduced by the $K^{*}$ polarization sum of Eq. \eqref{eq:pol} result fictitious once the value of $q^0$ at the poles is substituted in Eq. \eqref{eq:fnum} in the Wick rotation leading to Eq. \eqref{eq:Gtilde3}.

Taking into account that the $\eta(1405)$ is an $I=0$ object and that $K^{+}K^{-}$ and $K^{0}\bar{K}^{0}$ vertices appear with different sign, the amplitude of Eq. \ref{eq:scatt} is substituted now by
\begin{equation}
\label{eq:scattnew}
t_{f}=\tilde{G}(P,p_{\pi},m_{K^{+}},m_{K^{*+}})\ t_{K^{+}K^{-},f}-\tilde{G}(P,p_{\pi},m_{K^{0}},m_{K^{*0}})\ t_{K^{0}\bar{K}^{0},f}\ ,
\end{equation}
where $f$ now stands for $\pi^{+}\pi^{-}$ or $\pi^{0}\eta$, as before.
%************************************************************************************************************************************************
\subsection{Results with the triangular diagram}
\begin{figure}[ht]
\begin{minipage}[b]{0.7\linewidth}
\centering
\includegraphics[width=\textwidth]{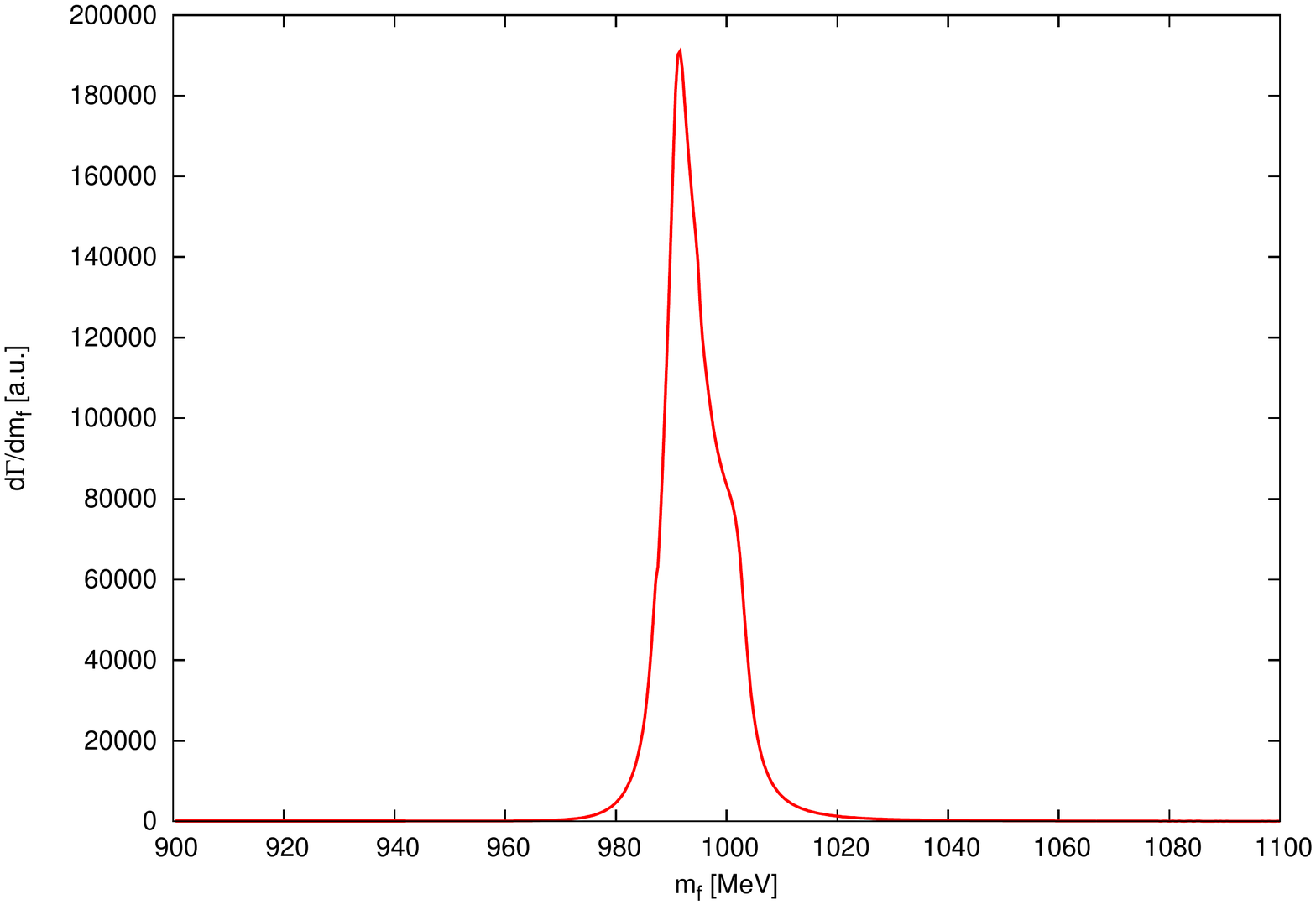}
\caption{$\frac{d\Gamma}{dm_f}$ for $\eta'\rightarrow\pi^{0}\pi^{+}\pi^{-}$ decay in the 
$f_0(980)$ region.}
\label{fig:f0trian}
\end{minipage}
\hspace{0.5cm}
\begin{minipage}[b]{0.7\linewidth}
\centering
\includegraphics[width=\textwidth]{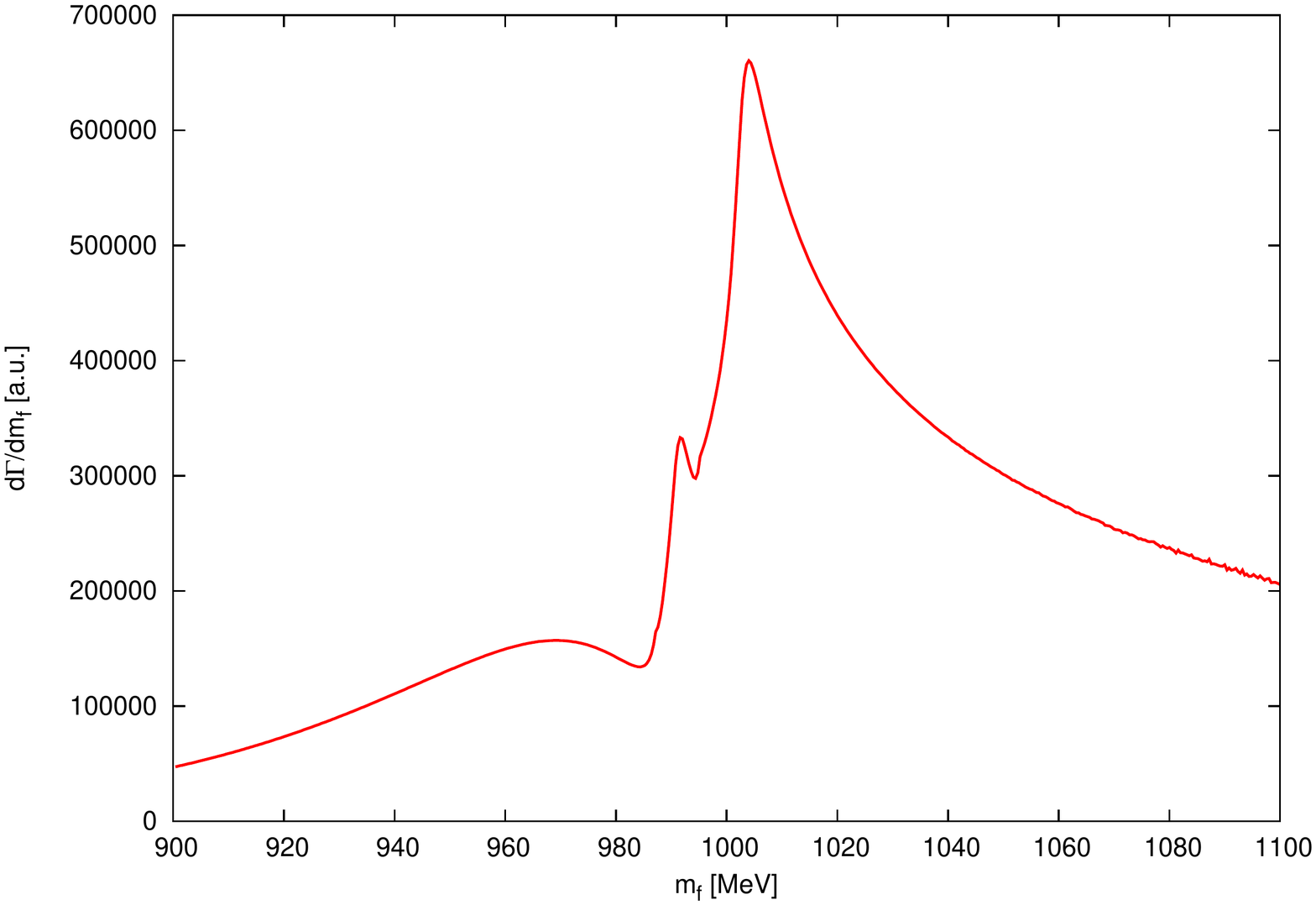}
\caption{$\frac{d\Gamma}{dm_f}$ for $\eta'\rightarrow\pi^{0}\pi^{0}\eta$ decay in the 
$a_0(980)$ region.}
\label{fig:a0trian}
\end{minipage}
\end{figure}
In Fig. \ref{fig:f0trian} we show the  result for $d\Gamma/dm_f$ for $\eta(1405)\rightarrow \pi^{0}\pi^{+}\pi^{-}$ and in Fig. \ref{fig:a0trian} for $\eta(1405)\rightarrow \pi^{0}\pi^{o}\eta$. 
\begin{figure}[ht]
\begin{minipage}[b]{0.7\linewidth}
\centering
\includegraphics[width=\textwidth]{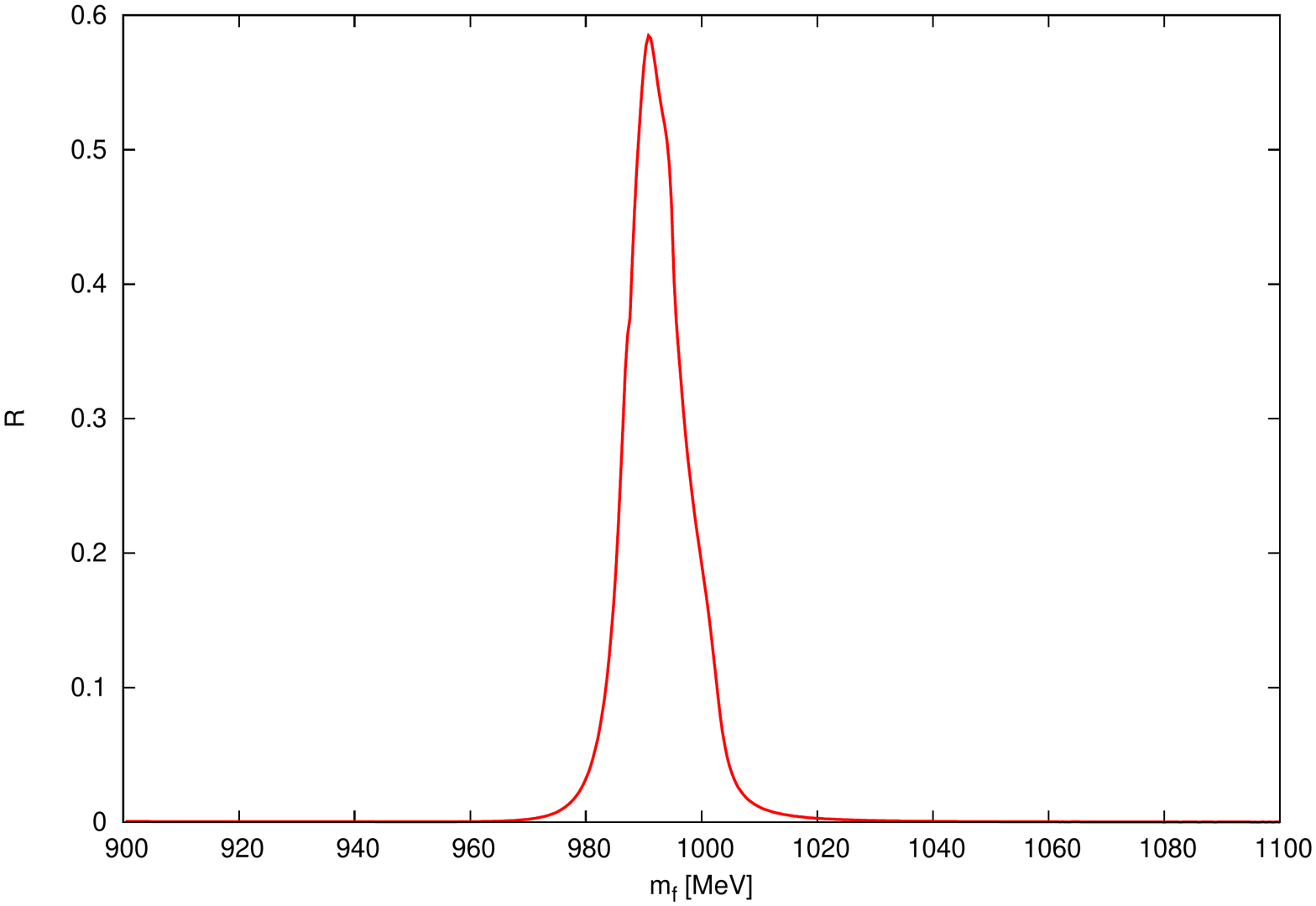}
\caption{Ratio $\left(\frac{d\Gamma}{dm_f}\right)_{\pi^{+}\pi^{-}}/\left(\frac{d\Gamma}{dm_f}\right)_{\pi^{0}\eta}$ as a function of $m_f$.}
\label{fig:ratiotrian}
\end{minipage}
\end{figure}
The shapes are similar to those in Figs. \ref{fig:figure2a} and \ref{fig:figure2b}, and however we can already observe that the ratio, depicted in Fig. \ref{fig:ratiotrian} is much bigger than that of Fig. \ref{fig:ratio1}. About a factor nine bigger. From these spectra we find that the ratio of integrated decay widths is now
\begin{equation}
\label{eq:ratttt}
\frac{\Gamma(\pi^{0},\pi^{+}\pi^{-})}{\Gamma(\pi^{0},\pi^{0}\eta)}\simeq 0.13\ .
\end{equation}

This $13\%$ is much closer to the experimental value of $(17.9  \pm 4.2)\%$, which has a lower 
limit of $13.7 \%$. Assuming similar theoretical uncertainties the results are 
compatible. We have made some estimates of the errors by changing the cutoff $q_{max}$ by $\pm 20\ MeV$, which moves the $f_{0}(980)$ and $a_{0}(980)$ peak in $\pi\pi$ and $\pi\eta$ scattering by about $8\ MeV$. We find that this change induces changes in the ratio of Eq. \eqref{eq:ratttt} by $0.01$. However, an uncertainty of $0.02$ is more indicated to account also for the uncertainties in the background subtraction. So we would be obtaining $(0.13\pm 0.02)$ for the fraction of decay rates.

This increase by about one order of magnitude with respect to the standard 
calculation is a consequence of the two neighboring singularities in the triangle diagram 
which is peculiar to the $\eta(1405)$ case.

We can now estimate the effect of having also $\pi^{0}\pi^{0}\eta$ in the primary production process. A triangular diagram of the type used for $\pi^{0}K\bar{K}$ production is not possible now. Indeed, one would have to substitute the $K^{*}$ by a $\rho$, but this is dynamically forbidden (no $\rho\pi^{0}\eta$ coupling). Then we must rely upon a contact term like assumed in the former section.

By recalling the exercise done in the former section (Eqs. \eqref{eq:RRexp}-\eqref{eq:RRexp3}) and 
the conclusion that positive values of R (with respect to an equivalent local $\pi^0 K 
\bar K$ production mechanism) were preferred, the inspection  of Fig. \eqref{eq:G} 
can give us a qualitative estimate of what adding this new primary $\pi^0 \pi^0 \eta$ 
production vertex can do to the widths, which is a moderate increase of the ratio 
$\Gamma(\pi^0, \pi^+ \pi^-)/\Gamma(\pi^0, \pi^0 \eta)$ by about $26\%$. This would 
provide a ratio around $16.4\%$ with an uncertainty of $2.5\%$, or rounding errors, a ratio of $(0.16\pm 0.03)$, in good agreement with the experimental values.

Now we come back to the BES experiment \cite{etabes}. In this experiment the authors cannot distinguish whether they have the $\eta(1405)$ or the $\eta(1475)$ resonance, so we must assume that they have a mixture of both. In order to account for this possibility, we have evaluated  the same ratio of rates as before assuming that we have now the $\eta(1475)$ resonance. The result that we obtain is
\begin{equation}
\label{eq:rat1475}
\frac{\Gamma(\pi^{0},\pi^{+}\pi^{-})}{\Gamma(\pi^{0},\pi^{0}\eta)}\Bigg|_{\eta(1475)}\simeq 0.16\ .
\end{equation}
This coincides with the centroid of our result of $(0.16\pm 0.03)$. We might also think about the possibility to have a contribution from the original $\pi\pi\eta$ channel. However, the same collaboration team reports for the mixture of the resonances in the $J/\psi\rightarrow\gamma\pi^{+}\pi^{-}\eta$,  $J/\psi\rightarrow\gamma K\bar{K}\eta$ a large dominance of the second process by nearly one order of magnitude \cite{baipi, baik}, which means we can neglet the primary $\pi\pi\eta$ channel in this case. Hence, assuming the same uncertainties as before, our final results for the $\eta(1405)$, the $\eta(1475)$, or a mixture of both, are given by
\begin{equation}
\label{eq:tutto}
\frac{\Gamma(\pi^{0},\pi^{+}\pi^{-})}{\Gamma(\pi^{0},\pi^{0}\eta)}= 0.16\pm 0.03\ .
\end{equation}

%************************************************************************************************************************************************
\section{Conclusions}

In this paper we have carried out a calculation of the decay rates of the
$\eta(1405) \to \pi^{0} f_0(980)(\pi^{+} \pi^{-})$ and $\eta(1405) \to \pi^{0} a_0(980)(\pi^{0} 
\eta)$ reactions with the aim of investigating the isospin violation in the first 
reaction which is tied to the $f_0(980)$-$a_0(980)$ mixing in the terminology of other 
works. We have abstained of talking about a measure of the mixing since in our formalism 
there is no transition of one to the other resonance but a simultaneous production of 
both once the problem is tackled with meson states in charge basis with different masses, 
where a small violation of isospin is immediately obtained. Since the two resonances are 
produced from the interaction of meson pairs, the process proceeds via a first step in 
which a $\pi^{0}$ and a pair of mesons are produced, and a second step in which the pair of 
mesons interacts. Isospin violation has then two sources, the first loop after the 
production, and the scattering matrices of meson-meson interaction. But in both cases the 
violation is tied to the difference of masses
between the charged and neutral kaons. This has as a consequence that the shape of the 
peak obtained for the $\pi^{+} \pi^{-}$ production in the first reaction has a very narrow 
width of the size of this mass difference, of the order of 9 MeV. This comes naturally in 
the approach and is in perfect agreement with the observation in the experiment.

In the first part we avoided making an explicit model for the reaction, but we assumed the primary production of $\pi^{0}PP$ to be given by a contact term and we could see that, invoking 
general principles and admitting large uncertainties in the input, we obtained a rate of 
$\pi^{+} \pi^{-}$ production versus $\pi^{0} \eta$ production which was rather small, of the 
order of one percent, which is in good agreement with the $f_0(980)$ and  
$a_0(980)$ mixing of the two BES experiments on $J/\psi \to \phi \pi \eta$ and $\chi_{c1} 
\to \pi^{0} \pi \pi$, with respect to the isospin allowed counterparts of $J/\psi \to \phi 
\pi \pi$ and $\chi_{c1} \to \pi^{0} \pi \eta$ \cite{besphiaf}. The rates obtained are also 
in agreement with those obtained in theoretical papers of the $J/\psi \to \phi \pi \eta$ 
versus
$J/\psi \to \phi \pi \pi$ \cite{ramonetchris,luisnew}. However, the rates obtained for 
the $\eta \to \pi^{0} \pi^{+} \pi^{-}$ versus $\eta \to \pi^{0} \pi^{0} \eta$
are very small compared to those claimed in the experiment \cite{etabes}, about one order of magnitude smaller. We tried to understand the situation by admitting a large admixture 
of I=1 in the $\eta(1405)$ wave function, but it required a very large I=1 component, not 
easily acceptable, and worse, it gave a signal for $f_0(980)$ production which had a 
width of the order of 20 MeV, which was much larger than the experimental one of the order 
of 9 MeV.

In the second part we followed the approach of \cite{wuzou} using the dominant primary 
production mechanism given by $\eta' \to K^* \bar K$ followed by $K^* \to K \pi$. The 
first loop now was quite different than for the contact interactions since the new 
singularity associated to $\eta' \to K^* \bar K$ played a very important role in the 
reaction. We found that using this new mechanism of production, the ratio of 
$\Gamma(\pi^0, \pi^+ \pi^-)/\Gamma(\pi^0, \pi^0 \eta)$ was increased by about one order 
of magnitude with respect to the results using the contact production vertices, providing 
results very close to those in the experiment. These results confirm the claims of 
\cite{wuzou}, where, however, a precise determination of that ratio could not be given 
since it was tied to unknown form factors needed to regularize the divergent loops. The 
use of the chiral unitary approach in the present work solved this problem since one 
could associate the regularizing cutoff in the new loops to the one used in meson-meson 
scattering to generate the $f_0(980)$ and $a_0(980)$ resonances dynamically. This allowed 
us to make quantitative predictions for the $\Gamma(\pi^0, \pi^+ \pi^-)/\Gamma(\pi^0, 
\pi^0 \eta)$ ratio, with a value $(0.16\pm 0.03)$, in basic agreement with experiment, of $(0.179\pm 0.04)$.

We also showed that the results obtained for that ratio were the same if we had the $\eta(1475)$ resonance instead of the $\eta(1405)$, or a mixture of the two, as seems to be the case in the BES experiment.

A final conclusion to be drawn is that the concept of  $f_0(980)-a_0(980)$ mixing is not 
very appropriate and different apparent ratios are obtained in different reactions. Then, the 
chiral unitary approach appears as an appropriate and accurate tool to use in order to 
analyze these reactions, and the present results, together with other results in 
different reactions on the mixing of these resonances, come to strengthen the support for 
the $f_0(980)$ and $a_0(980)$ resonances as dynamically generated from the meson-meson 
interaction.

\section*{Acknowledgments}

This work is partly supported by DGICYT contract FIS2011-28853-C02-01, the Generalitat 
Valenciana in the program Prometeo, 2009/090, and the EU Integrated Infrastructure 
Initiative Hadron Physics 3 Project under Grant Agreement no. 283286, the National Natural
Science Foundation of China under Grant 11035006, 11261130311 (CRC110 by DFG and NSFC), 
the Chinese Academy of Sciences under Project No.KJCX2-EW-N01 and the Ministry of Science
and Technology of China (2009CB825200), the National Natural
Science Foundation of China under Grant 11165005.

\end{document}